# First principles investigation of thermal conductivity in Magnesium Selenide(MgSe) with different crystalline phase

Rajmohan Muthaiah, Jivtesh Garg

School of Aerospace and Mechanical Engineering, University of Oklahoma, Norman, OK-73019, USA

**Abstract:** Magnesium Selenide (MgSe) is a wide bandgap semiconductor with applications in optoelectronics and energy conversion technologies. Understanding thermal conductivity ($k$) of MgSe is critical for optimum design of thermal transport in these applications. In this work, we report the temperature and length dependence lattice thermal conductivity of magnesium selenide (MgSe) with different crystallographic phases; zincblende, rocksalt, wurtzite and nickel arsenic, using first principles computations. Computations reveal significant differences in thermal conductivity ($k$) of MgSe for different phases. The observed trend in thermal conductivities is : $k_{NiAs} < k_{rocksalt} < k_{wurtzite} < k_{zincblende}$. Our first principles calculations show a room temperature low $k$ of 4.5 Wm$^{-1}$K$^{-1}$ for the NiAs phase and a high $k$ of 20.4 W/mK for wurtzite phase. These differences are explained in terms of a phonon band gap in the vibrational spectra of zincblende and wurtzite phases, which suppresses scattering of acoustic phonons, leading to high phonon lifetimes.

**Keywords:** Density functional theory, magnesium, selenium, thermal conductivity, semiconductors

**Introduction:** Wide bandgap materials have attracted scientific and technological interest due to several unique properties such as reduced energy consumption, low power loss, ability to accommodate high operating temperatures, high switching speed and high frequencies[1-3] and good thermoelectric energy conversion[4-7]. Magnesium chalcogenides such as magnesium sulphide (MgS), magnesium selenide (MgSe) and magnesium telluride (MgTe) are wide bandgap semiconductors which are extensively studied for their electronic[8-10], magnetic[9], optical[8], structural[9, 11-13] and vibrational[14, 15] properties. Understanding thermal conductivity of these materials is critical for optimum thermal design of devices based on these materials. There are, however, limited studies on thermal properties which is critical for wide range of applications such as thermoelectrics[16-21], thermal management systems[22-27], opto-electronics[28], thermal barrier coatings[29-31] and solar cells[32-34]. In this work, we report the temperature and size dependent thermal conductivity of magnesium selenide (MgSe) with different crystalline phases using first principles calculations. MgSe exists in four crystalline phases; zincblende(ZB), rocksalt(RS), wurtzite(WZ) and nickel arsenic(NiAs)[14]. At 300 K, the first principles computed thermal conductivities of MgSe are – a) 4.54 Wm$^{-1}$K$^{-1}$ along a-axis and 6.37 Wm$^{-1}$K$^{-1}$ along c-axis for NiAs structure, b) 11.89 Wm$^{-1}$K$^{-1}$ for Rocksalt structure, c) 19.58 Wm$^{-1}$K$^{-1}$ along a-axis and 20.39 Wm$^{-1}$K$^{-1}$ along c-axis for WZ structure and d) 21.27 Wm$^{-1}$K$^{-1}$ for Zinc-Blende structure. Understanding of differences in thermal conductivity is achieved through analysis of differences in phonon scattering arising from different phonon dispersions for different structures.

**Computational Methods:**

All the first principles calculations were performed using QUANTUM ESPRESSO[35] package. Norm-conserving pseudopotentials with local density (LDA)[36] exchange-correlation functional are used for electronic calculations. The geometries of the zinc-blende and rocksalt MgSe with 2 atom unit cell and wurtzite and NiAs structures with 4 atom unit cell, were optimized until forces on all atoms were less than 10$^{-6}$ Ry/bohr. Plane-wave energy cutoff of 70 Ry was used for electronic calculations. Monkhorst-Pack[37] $k$-point mesh sizes of 8 x 8 x 8 and 12 x 12 x 8 were used for zinc-blende/rocksalt and wurtzite/NiAs structures, respectively, to integrate over the Brillouin zone.

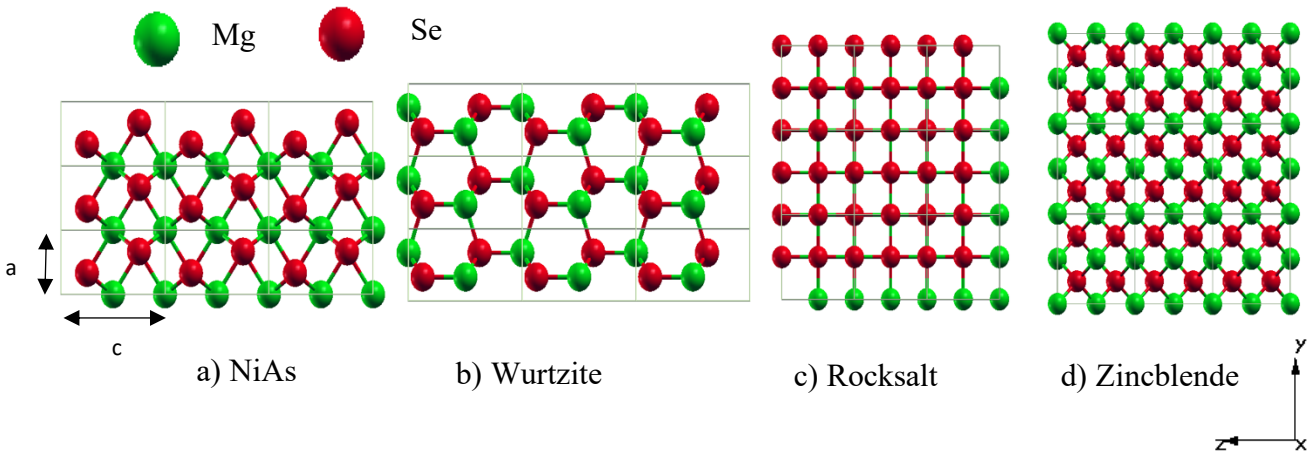

Figure 1a-d): Crystal structure of MgSe with crystalline phases; NiAs (a=7.216 bohr, c/a=1.6672), wurtzite (a=7.924 bohr, c/a=1.6149), rocksalt (a=10.2617 bohr) and zincblende (a=11.16 bohr) respectively.

Relaxed structures with equilibrium lattice constants of MgSe with different lattice crystal phases are shown in Fig 1 and also listed in Table 1 (in excellent agreement with previously published values[10, 14, 38-40]).

Lattice thermal conductivity ($k$) was computed by solving phonon Boltzmann transport equation (PBTE)[41] in both single mode relaxation approximation (SMRT)[42] and exactly by using a variational method. Expression for thermal conductivity ($k$) obtained by solving PBTE in the single mode relaxation time (SMRT) approximation[43] is given by,

$$k_\alpha = \frac{\hbar^2}{N\Omega k_b T^2} \sum_\lambda v_{\alpha\lambda}^2 \omega_\lambda^2 \bar{n}_\lambda (\bar{n}_\lambda + 1)\tau_\lambda \quad (1)$$

where, $\alpha$, $\hbar$, N, $\Omega$, $k_b$, T, are the cartesian direction, Planck constant, size of the **q** mesh, unit cell volume, Boltzmann constant, and absolute temperature respectively. $\lambda$ represents the vibrational mode (**q**j) (**q** is the wave vector and j represent phonon polarization). $\omega_\lambda$, $\bar{n}_\lambda$, and $v_{\alpha\lambda}$ ($= \partial\omega_\lambda/\partial q$) are the phonon frequency, equilibrium Bose-Einstein population and group velocity along cartesian direction $\alpha$, respectively of a phonon mode $\lambda$. $\omega_\lambda$, $\bar{n}_\lambda$, and $c_{\alpha\lambda}$ are derived from the knowledge of phonon dispersion computed using 2$^{nd}$ order IFCs. $\tau_\lambda$ is the phonon lifetime and is computed using the equation,

$$\frac{1}{\tau_\lambda} = \pi \sum_{\lambda'\lambda''} |V_3(-\lambda, \lambda', \lambda'')|^2 \times [2(n_{\lambda'} - n_{\lambda''})\delta(\omega(\lambda) + \omega(\lambda') - \omega(\lambda'')) + (1 + n_{\lambda'} + n_{\lambda''})\delta(\omega(\lambda) - \omega(\lambda') - \omega(\lambda''))]$$

(2)

where, $\frac{1}{\tau_\lambda}$ is the anharmonic scattering rate based on the lowest order three phonon interactions and $V_3(-\lambda, \lambda', \lambda'')$ are the three-phonon coupling matrix elements computed using both harmonic (2$^{nd}$ order) and anharmonic (3$^{rd}$ order) interatomic force constants. 2$^{nd}$ and 3$^{rd}$ order interatomic force constants were derived from density-functional perturbation theory (DFPT)[44, 45]. Harmonic force constants were computed on an 8 x 8 x 8 **q**-grid for ZB and RS systems and on a 9 x 9 x 6 grid for WZ and NiAs structures. Anharmonic force constants were computed on a 4 x 4 x 4 grid for ZB and RS and on a 3 x 3 x 2 grid for WZ and NiAs structures, using D3Q[41, 46, 47] package within QUANTUM-ESPRESSO. Acoustic sum rules were imposed on both harmonic and anharmonic interatomic force constants. Phonon linewidth and lattice thermal conductivity were calculated using 'thermal2' package within QUANTUM ESPRESSO. For these calculations, *q*-mesh of 30 x 30 x 30 was used for ZB and RS structures, while a mesh of 30 x 30 x 20 was used for WZ and NiAs structures. Iterations in the exact solution of the PBTE were performed until Δ***k*** between consecutive iterations diminished to below $1.0e^{-5}$. *k* values were typically converged after 4 iterations. Casimir scattering[48] was imposed to include the effect of boundary scattering for computing length dependent thermal conductivity in the nanoscales. Phonon-isotope scattering is included for the effect of isotope variation with naturally occurring isotopes of Mg and Se[49]. Elastic constants were computed using QUANTUM ESPRESSO thermo_pw package. Voigt-Reuss-Hill approximation[50] was used to calculate Bulk modulus, Shear modulus(G), Young's Modulus(E) and Poisson's ratio (υ).

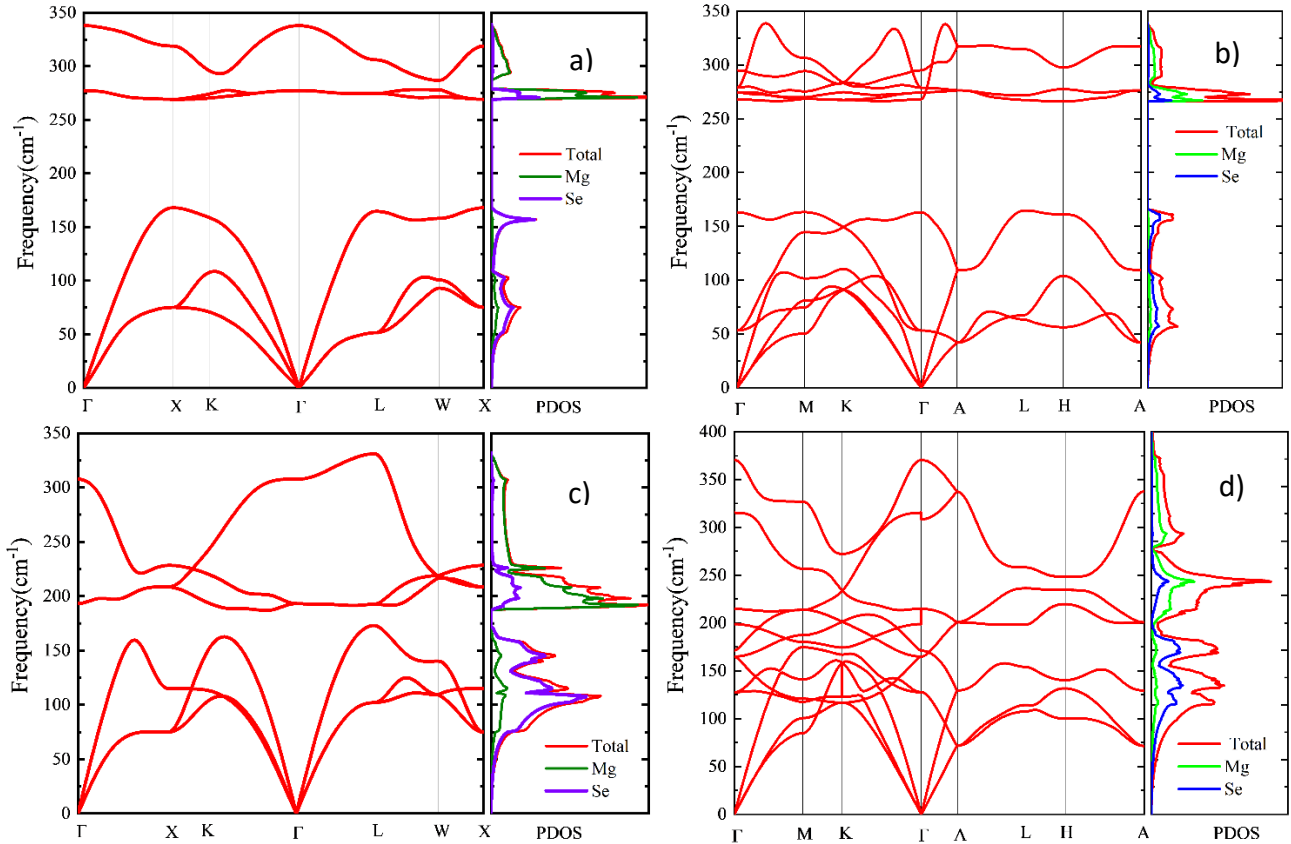

Figure 2: Phonon dispersion and phonon density of states(PDOS) of MgSe with crystalline phase; a) zincblende b) wurtzite c) rocksalt and d) nickel arsenic.

**Results and Discussion:** Phonon dispersion and phonon density of states (PDOS) for the MgSe with crystalline phases of zincblende(ZB), wurtzite (WZ), rocksalt(RS) and nickel arsenic(NiAs) are shown in Fig 2 a-d. Computed dispersions are in good agreement with previously reported values[14]. Elastic properties such as Young's modulus (E), Bulk modulus (B), Shear modulus (G) and Poisson's ratio based on Voigt-Ruess-Hill approximation are listed in Table 1 and are also in excellent agreement with the previously published work[14].

**Table 1: Lattice constants, Bulk modulus(B), Youngs modulus(E), Shear modulus(G) and poisson's($\upsilon$) ratio of MgSe with different crystal phase.**

| S. No | Crystal phase | a (bohr) | c/a | B (GPa) | E(GPa) | G(GPa) | $\upsilon$ |
|---|---|---|---|---|---|---|---|
| 1. | Nickel arsenic (NiAs) | 7.216 | 1.667 | 67.36 | 92.8 | 36.53 | 0.2703 |
| 2. | Wurtzite (WZ) | 7.9238 | 1.615 | 50.7 | 55.34 | 21 | 0.3182 |
| 3. | Rocksalt (RS) | 10.2617 | | 67.7 | 113.07 | 46.28 | 0.2216 |
| 4. | Zincblende (ZB) | 11.16 | | 49.695 | 47.903 | 17.933 | 0.3356 |

Temperature dependent lattice thermal conductivity ($k$) of MgSe for different crystalline phases is shown in Fig 3a. Single-mode relaxation results (SMA) are 5% less than the iterative solution. At room temperature (300K), computed $k$ of pure MgSe is as follows: $k_{NiAs}$ (4.54 Wm$^{-1}$K$^{-1}$ along a-axis and 6.37 Wm$^{-1}$K$^{-1}$ along c-axis) < $k_{RS}$ (11.89 Wm$^{-1}$K$^{-1}$) < $k_{WZ}$ (19.58 Wm$^{-1}$K$^{-1}$ along a-axis and 20.39 Wm-1K-1 along c-axis) < $k_{ZB}$ (21.27 Wm$^{-1}$K$^{-1}$). Fig 3b shows the length dependent thermal conductivity of MgSe between 10 nm and 1μm. At 300K and at 100 nm, $k$ of different crystalline phases are as follows: $k_{NiAs}$ (3.25 Wm$^{-1}$K$^{-1}$ along a-axis and 4.67 Wm$^{-1}$K$^{-1}$ along c-axis) < $k_{RS}$(5.71 Wm$^{-1}$K$^{-1}$) < $k_{WZ}$ (8.05 Wm$^{-1}$K$^{-1}$ a-axis and 8.76 Wm$^{-1}$K$^{-1}$ along c-axis) < $k_{ZB}$(8.82 Wm$^{-1}$K$^{-1}$).

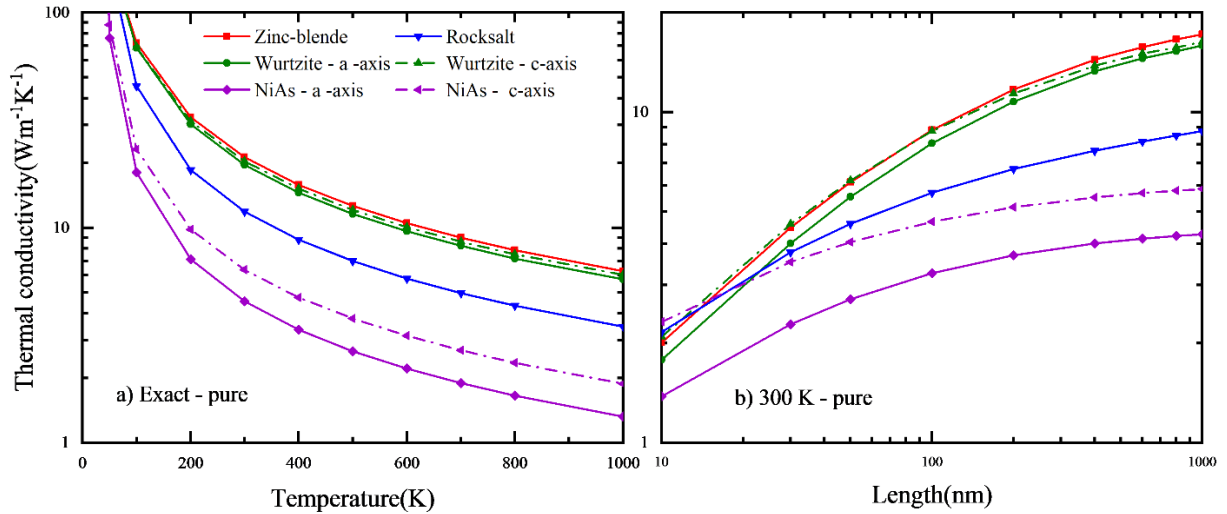

Figure 3: Temperature dependent lattice thermal conductivity of isotopically pure MgSe with different crystalline phase with a) iterative solution of BTE b) Length dependent thermal conductivity of MgSe at room temperature(300K).

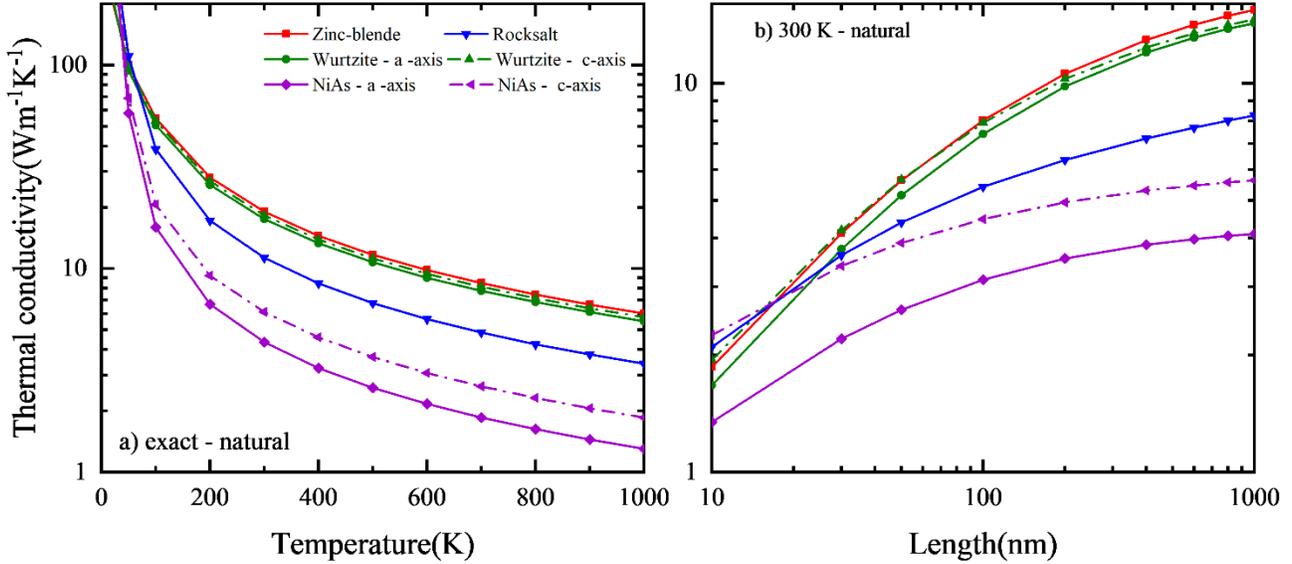

Figure 4a and b) Temperature and length dependent (300 K) thermal conductivity of natural MgSe for different crystalline phase.

Lattice thermal conductivity of naturally occurring MgSe, which includes the effect of isotopic disorder is shown in Fig 4. Thermal conductivity of naturally occurring MgSe at 300K is as follows: $k_{NiAs}$ (4.36 Wm$^{-1}$K$^{-1}$ along a-axis and 6.13 Wm$^{-1}$K$^{-1}$ along c-axis) < $k_{RS}$(11.31 Wm$^{-1}$K$^{-1}$) < $k_{WZ}$(17.5 Wm$^{-1}$K$^{-1}$ a-axis and 18.3 Wm-1K-1 along c-axis) < $k_{ZB}$(19.04 Wm$^{-1}$K$^{-1}$). These values show that isotopic scattering reduces lattice thermal conductivity by a maximum of ~10% for ZB and WZ phase.

As seen above, thermal conductivity of the NiAs crystal phase is the lowest, while that of zincblende phase is the highest. Thermal conductivity of MgSe with zincblende crystal structure is 4.68 times that of the NiAs phase along a-axis at 300 K. In Fig. 5, we also compare contributions of different vibrational modes to overall thermal conductivity in different crystalline phases. Interestingly, in NiAs and wurtzite MgSe, $k$ contribution of optical phonons is significantly higher than for the case of zincblende and rocksalt phases.

The higher thermal conductivity of zincblende phase is due to the suppression of three-phonon scattering mediated by a large phonon bandgap (~100 cm$^{-1}$) in the phonon dispersion of zinc-blende structure. We have presented the phonon scattering rates (inverse of phonon lifetime) and phonon group velocities of MgSe with different crystalline phases in Fig 6. We can observe from Figs 6a and d that, scattering rate of transverse acoustic modes, $TA_1$ and $TA_2$, for NiAs is approximately one order of magnitude higher than that of the zincblende phase. This causes a dramatic reduction in thermal conductivity contributions of $TA_1$ and $TA_2$ phonon modes in NiAs structure, reducing the overall thermal conductivity.

The above difference in scattering rates can be understood by observing that anharmonic scattering of phonons through the lowest-order three phonon processes can be classified into two categories - absorption scattering process, where a phonon mode $(q\omega)$ scatters by absorbing another phonon mode $(q'\omega')$, yielding a higher energy $(q''\omega'')$ phonon mode, and decay processes, where a phonon mode decays into two lower energy phonons. These processes satisfy energy and momentum conservation given by, $\omega+\omega'=\omega''$ (energy), $q + q' = q''$ (momentum) for absorption process and $\omega=\omega'+\omega''$ (energy), $q=q'+q''$ (momentum) for decay process.

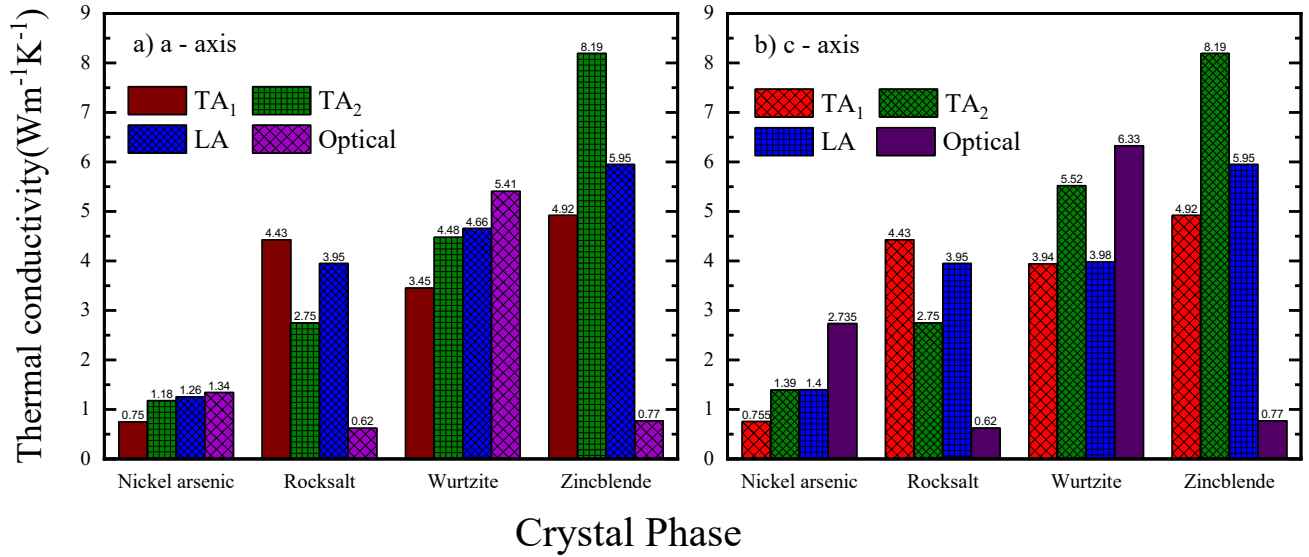

Figure 5: Mode contribution thermal conductivity of MgSe at 300K for different crystalline phase.

The large energy gap in the phonon dispersion of zincblende structure suppresses the absorption scattering channels for acoustic phonons involving scattering of an acoustic phonon by absorbing another acoustic phonon to convert into an optical phonon. The large energy gap in the phonon dispersion of the zinc-blende structure prohibits energy conservation ($\omega+\omega'=\omega''$) for such absorption scattering channels. This is seen through an example, where, an acoustic phonon of frequency 100 cm$^{-1}$ cannot scatter into an optical phonon, even by absorbing the highest frequency

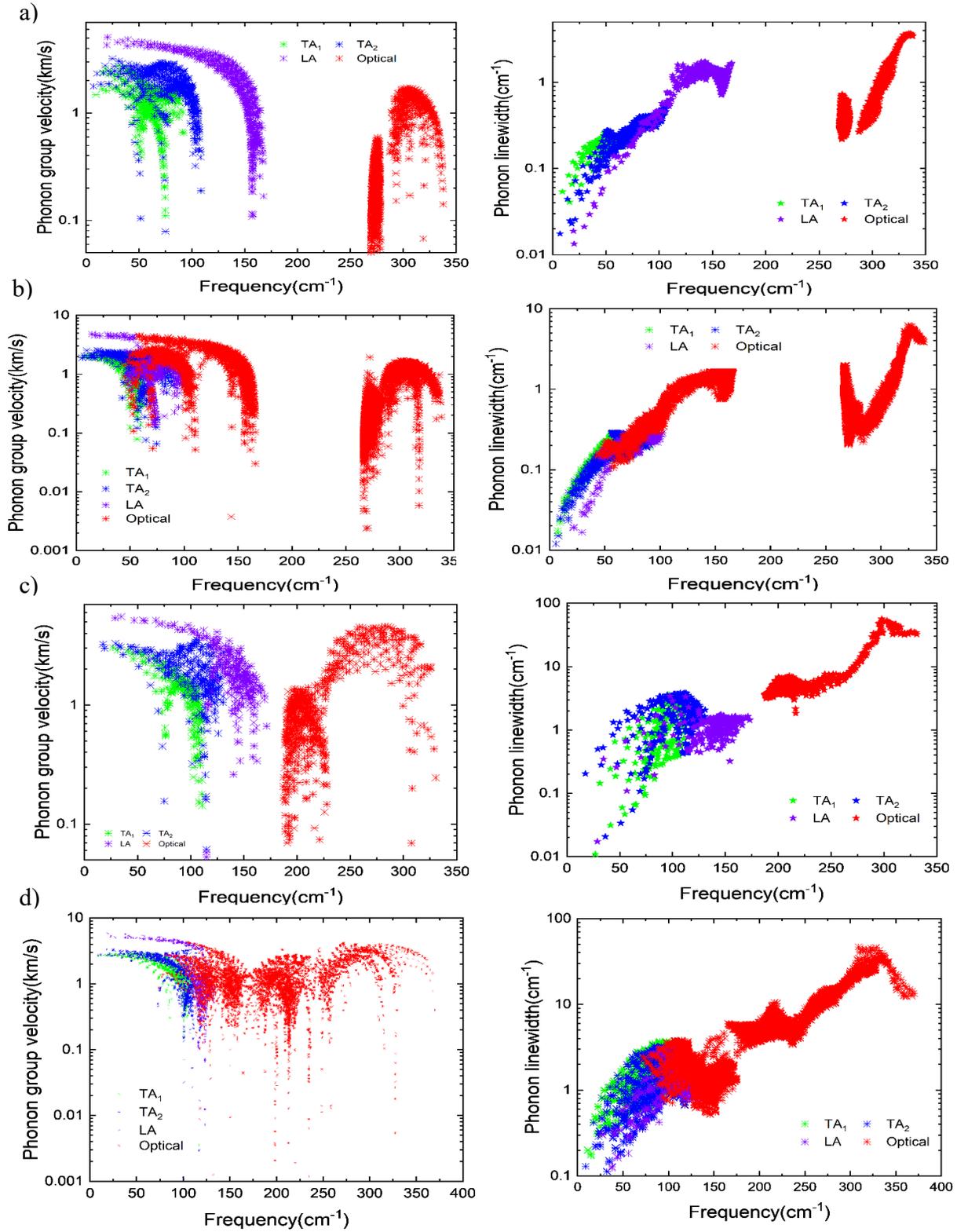

Figure 6: Phonon group velocity and scattering of MgSe with crystalline phase; a) zincblende b) wurtzite c) rocksalt and d) nickel arsenic

acoustic phonon (168.05 cm$^{-1}$). This is because the lowest optical phonon frequency 269.25 cm$^{-1}$

is higher than the sum of the frequencies of above two listed acoustic phonons. This elimination of absorption scattering channels in zinc-blende structure dramatically decreases overall scattering rates in zinc-blende case. We have shown this for all crystalline phases in Fig. 7, where we compare the magnitude of absorption scattering channel with the overall scattering rates.

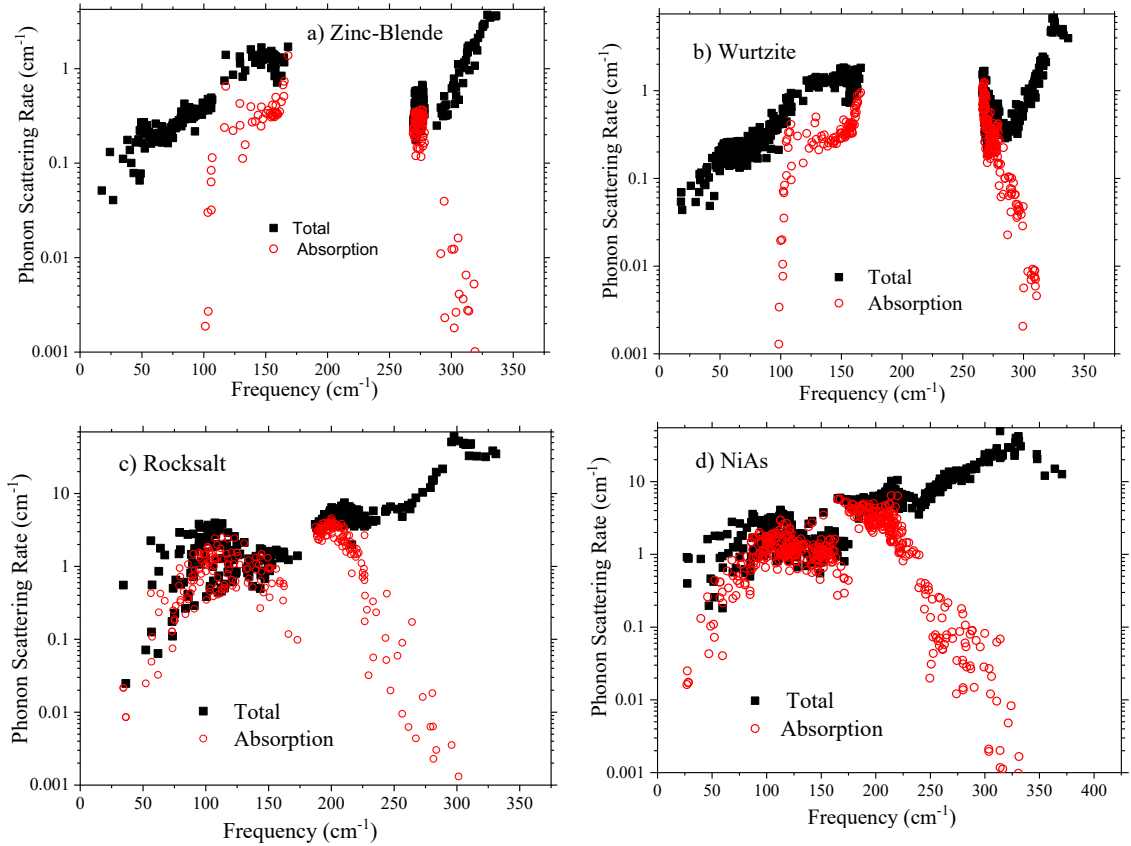

Figure 7: Role of absorption scattering channels in overall phonon scattering of MgSe with crystalline phase; a)zincblende b) wurtzite c) rocksalt and d) nickel arsenic

It can be seen that for zincblende and wurtzite structures (with a large bandgap in their phonon dispersions), the magnitude of absorption channels is significantly smaller than for the case of rocksalt and NiAs structures. Below a frequency of 100 cm$^{-1}$, the absorption channel is seen to be almost completely absent in zinc-blende and wurtzite structures. This smaller rate of absorption scattering in zincblende and wurtzite structures also leads to a smaller overall scattering rate in these structures. At a frequency of 50 cm$^{-1}$, the overall scattering rate in zinc-blende and wurtzite structures is ~0.1 cm$^{-1}$, almost an order of magnitude lower, relative to the scattering rate of 1 cm$^-$ [1] in NiAs structure. The higher scattering rates in rocksalt and NiAs structures (due to smaller or absent phonon band gap) lead to lower thermal conductivity in these crystalline phases.

Higher contribution of optical phonon modes to overall thermal conductivity in NiAs and Wurtzite structures can now be understood in terms of the phonon band gap in these materials. In NiAs structure, the large scattering rates of acoustic phonons (due to absence of a band gap in phonon dispersion), imply that the scattering rates of low frequency optical phonons become comparable to that of acoustic phonons. Significant group velocities of optical phonons in NiAs structure combined with comparable phonon scattering rates to acoustic phonons, leads to high thermal conductivity contribution of optical phonons in NiAs structure. In Wurtzite crystalline phase, the high thermal conductivity of optical phonons arises due to the large phonon band gap in the dispersion. Fig. 2b shows that some of the optical phonons are below the band gap. Similar to the case of acoustic phonons, these optical phonons also experience inhibited scattering from optical phonons above the band gap, resulting in an increase in their lifetimes. These higher lifetimes result in a higher contribution of optical phonons to overall thermal conductivity in wurtzite structure.

**Conclusion:** In this work, thermal conductivities of magnesium selenide (MgSe) with four crystalline phases; zincblende, rocksalt, wurtzite and nickel arsenic were computed through first principles calculations. Computations reveal significant differences in thermal conductivities of different phases; at 300 K, lowest $k$ value of 4.54 W/mK (along a-axis) is obtained for NiAs phase, while highest value of 21.27 W/mK is obtained for zincblende phase. These differences are found to be due to a large energy gap in the phonon dispersion of zincblende and wurtzite structures, which suppresses scattering of acoustic phonons by optical phonons, leading to longer phonon lifetimes and higher thermal conductivity in zincblende and wurtzite phases. Isotopic disorder scattering has minimal effect (less than 10%) on overall thermal conductivity. First principles calculations further reveal NiAs and wurtzite phases to have significant contributions from optical phonons. This effect is again explained in terms of the phonon band gap. At nanometer length scales such as 100 nm, thermal conductivity of less than 3.25 $Wm^{-1}K^{-1}$ for MgSe with NiAs crystalline phase, shows a promising application of MgSe for thermoelectric technologies.

**Acknowledgements:** R.M and J.G would like to acknowledge OU Supercomputing Center for Education Research (OSCER) for providing computational resources. J.G and R.M acknowledge financial support from NSF CAREER grant, Award # 1847129.